\documentclass[12pt]{iopart}

\usepackage{amssymb}
\usepackage{envmath}
\usepackage{graphicx}

\begin{document}

\newcommand{\Mn}{{\mathcal{M}_n}}
\newcommand{\Jn}{{\mathcal{J}_n}}
\newcommand{\q}{{\widetilde{q}}}
\newcommand{\D}{{\widetilde{\mathcal{D}}}}
\newcommand{\K}{{\widetilde{K}}}
\newcommand{\R}{{\widetilde{\mathcal{R}}}}
\newcommand{\St}{{\mathcal{S}}}
\newcommand{\twoeps}{{\widetilde{\epsilon}}}

\newcommand{\M}[1]{{\mathcal{M}_{#1}}}
\newcommand{\J}[1]{{\mathcal{J}_{#1}}}

\title{Tidal deformations of spinning black holes in Bowen-York initial data} 

\author{Miriam Cabero$^{1,2}$ and Badri Krishnan$^{1}$} 

\address{$^1$Max Planck Institute for Gravitational Physics, Albert
  Einstein Institute, Callinstrasse 38, 30167 Hannover, Germany }

\address{$^2$Leibniz Universit\"at Hannover, Appelstrasse 2, 30167
  Hannover, Germany}

\ead{badri.krishnan@aei.mpg.de}

\begin{abstract}
  We study the tidal deformations of the shape of a spinning black
  hole horizon due to a binary companion in the Bowen-York initial
  data set.  We use the framework of quasi-local horizons and identify
  a black hole by marginally outer trapped surfaces.  The intrinsic
  horizon geometry is specified by a set of mass and
  angular-momentum multipole moments $\Mn$ and $\Jn$ respectively.
  The tidal deformations are described by the change in these
  multipole moments caused by an external perturbation. This leads us
  to define two sets of dimensionless numbers, the tidal coefficients
  for $\Mn$ and $\Jn$, which specify the deformations of a black hole
  with a binary companion.  We compute these tidal coefficients in a
  specific model problem, namely the Bowen-York initial data set for
  binary black holes.  We restrict ourselves to axisymmetric
  situations and to small spins.  Within this approximation, we
  analytically compute the conformal factor, the location of the
  marginally trapped surfaces, and finally the multipole moments and
  the tidal coefficients.
\end{abstract}


\section{Introduction}
\label{sec:intro}

Among the fundamental properties of any classical or quantum
mechanical physical system is its response to external perturbations.
For example, the study of elasticity is the study of the deformation
of a solid body subject to an external force; in quantum mechanics, an
important property of atoms is the splitting of atomic spectral lines
in the presence of external electric and magnetic fields.  In
gravitational physics, an important example is the deformation of a
star due to the gravitational field of a binary companion.  This paper
studies the deformation of a black hole horizon subject to an external
perturbation.  One of the earliest discussions of tidal deformations
in general relativity, the Love numbers, and their role in formulating
the laws of motion is due to Damour \cite{Damour:1983}.  Some recent
studies of tidal deformations, Love numbers for neutron stars and
their potential implications for gravitational wave observations, are
\cite{Hinderer:2007mb,Flanagan:2007ix,Damour:2009vw,Damour:2012yf,Yagi:2013awa}.
Love numbers for non-spinning black holes are discussed in
\cite{Binnington:2009bb}.  

Both in Newtonian gravity and in general relativity, one could
consider either i) the deformations in the gravitational field of the
object at large distances from it, or ii) the change in the shape of
the body itself.  However, the relationship between the two
calculations is yet to be fully understood in general relativity.  The
papers cited in the previous paragraph were, for the most part,
concerned with the distortions of the body's asymptotic gravitational
field.  Deformations of the shape of black hole horizons (again in the
non-spinning case) are discussed in
\cite{Damour:2009va,Landry:2014jka}.  In this paper we shall consider,
for the first time to our knowledge, the deformation of a spinning
black hole and in particular, the deformation of its horizon shape.
We shall assume that the black hole angular momentum\footnote{In this
  paper by ``angular momentum'' we shall always mean the intrinsic
  angular momentum of the black hole.} is small and that the companion
is far away (compared to the mass of the black hole).  Furthermore, we
shall specialize to the manifestly axisymmetric case when the black
hole angular momentum and the separation vector between the black
holes are parallel to each other.

The essential ingredients in our calculation are the invariant horizon
multipole moments $\Mn$ and $\Jn$ for $n=0,1,2,\ldots$. These moments
fully characterize the intrinsic geometry of a black hole horizon and
will be affected by the external field.  We shall therefore begin with
a brief introduction to these moments.  Our goal will be to compute
how these moments are affected when a binary companion is introduced.
We shall work with a particular model binary system, namely black
holes in the Bowen-York initial data set \cite{Bowen:1980yu}.  This is
one of the simplest ways of studying a binary black hole system
consisting of spinning components.  The Bowen-York initial data
construction assumes that the spatial 3-metric is conformally flat and
it provides a prescription for solving the Hamiltonian and momentum
constraints for an arbitrary number of black holes including both
angular momentum $\mathbf{J}$ and linear momentum $\mathbf{P}$ for
each black hole.  We shall solve for the conformal factor
perturbatively assuming that both $\mathbf{J}$ and $\mathbf{P}$ are
small in magnitude.  This allows us to find the location of the
marginally trapped surfaces perturbatively and to thereby calculate
the black hole source multipole moments.  We can then identify how the
multipole moments are affected by the presence of the second black
hole and therefore find a set of numbers which uniquely characterize
how the moments are affected by an external perturbation.  The
calculation of these coefficients is the main result of this paper.

It is important to keep in mind an important caveat here.  From a
physical viewpoint, what we really want is to carry out a similar
computation for two Kerr black holes rather than for Bowen-York black
holes.  It is known that the Kerr spacetime does not admit conformally
flat spatial slices \cite{Garat:2000pn} and thus, the Bowen-York black
hole horizon is expected \emph{apriori} to be different from a Kerr
horizon.  In binary black hole numerical simulations which start with
Bowen-York data, it is found that the initial deviations from Kerr are
radiated away in the so called ``junk radiation'' and the individual
black hole horizons very quickly become indistinguishable from a Kerr
horizon.  While this is not a problem for the numerical simulations
which can ignore the initial burst of junk radiation, in our case this
will be more important.  The tidal properties of a Bowen-York black
hole may well be quite different from a Kerr black hole.  While it
would be interesting to investigate this further, it is nevertheless
useful since this would be the first such calculation for a spinning
black hole.  It also provides an interesting application of the
horizon multipole moments which clearly quantify the deviations of a
Bowen-York black hole from the Kerr horizon.  Since the Bowen-York
data set is commonly used as a starting point for numerical relativity
calculations, this might be useful for numerical relativity
applications.  There are numerous suggestions for constructing initial
data which resembles a system of two Kerr black holes more closely
(see e.g. \cite{Dain:2000hk}), but the Bowen-York data is the simplest
example for a spinning black hole.

A similar comment applies in fact also to non-spinning black holes.
In principle we would like to consider a spacetime consisting of two
Schwarzschild black holes far away from each other and falling head-on
towards each other.  However, there are potentially different ways of
approximating this physical situation.  We could, as in
\cite{Damour:2009va}, note that the spacetime must be axisymmetric and
thus model it by a static Weyl metric.  This will have the unphysical
feature that the black holes will continue to remain in a static
configuration.  The two black holes are held apart by a ``strut'' and
it is not clear how one should separate the influence of this strut
from the gravitational interaction between the two bodies.
Alternatively, we could consider the Brill-Lindquist
\cite{Brill:1963yv} or Misner \cite{Misner:1960zz} initial data sets
for binary systems, both of which contain two black holes initially at
rest.  None of these choices are perhaps entirely unreasonable
\footnote{Each of these are unphysical in their own way.  The Weyl
  metric approach yields a static solution as mentioned previously,
  the Brill-Lindquist data is time symmetric, and the Misner data
  represents a wormhole connecting the two black holes.  Thus,
  strictly speaking, none of these can represent two Schwarzschild
  black holes which were far away in the past and are coming closer to
  each other.} but we cannot suppose that they will yield the same
values of the horizon Love numbers or in general, the tidal
coefficients.

The plan for the rest of this paper is as follows.  In
Sec.~\ref{sec:multipoles} we shall review the definitions of the
horizon multipole moments and describe how these moments change under
the influence of an external perturbation.  The Bowen-York initial
data set is described briefly in Sec.~\ref{sec:bowenyork}.
Sec.~\ref{sec:singlebh} considers a single spinning Bowen-York black
hole.  Sec.~\ref{sec:binary} discusses a binary system and the
deformation of the black hole multipole moments due to a binary
companion. Finally Sec.~\ref{sec:conclusions} discusses some
implications of these results and directions for future work.  We
shall work in geometric units with $G=c=1$.  The spacetime metric,
with signature $(-+++)$, will be denoted by $g_{ab}$ and $\nabla_a$
will be the derivative operator compatible with it.  Our convention for the
Riemann tensor $R_{abcd}$ is
$(\nabla_a\nabla_b-\nabla_b\nabla_a)\omega_c = {R_{abc}}^d\omega_d$.

\section{The horizon multipole moments}
\label{sec:multipoles}

\subsection{The general framework}
\label{subsec:multipoles:general}

We shall use the framework of quasi-local horizons to describe black
holes.  This encompasses a wide range of physical situations: black
holes in equilibrium are modeled by isolated horizons
\cite{Ashtekar:1998sp,Ashtekar:2000sz,Ashtekar:1999yj,Ashtekar:2001jb,Ashtekar:2000hw,Booth:2001gx},
and a black hole growing due to in-falling matter/radiation is modeled
as a dynamical horizon \cite{Ashtekar:2002ag,Ashtekar:2003hk}.  Both
of these are closely related to the notion of trapping horizons
introduced earlier by Hayward
\cite{Hayward:1993wb,Hayward:1994yy,Hayward:2006ss,Hayward:2008ti}.
All these notions build on the idea of marginally outer trapped
surface.  Let $\St$ be a closed two-dimensional surface.  Let $\ell^a$
and $n^a$ be its outward and inwards pointing null normals
respectively.  $\mathcal{S}$ is said to be a marginally trapped
surface if the expansions of $\ell^a$ and $n^a$, denoted by
$\Theta_{(\ell)}$ and $\Theta_{(n)}$ respectively, satisfy
$\Theta_{(\ell)}=0$ and $\Theta_{(n)}<0$.  $S$ is said to be a
marginally \emph{outer} trapped surface if $\Theta_{(\ell)}=0$ with no
restriction on $\Theta_{(n)}$.  In practice we will not check the
condition on $\Theta_{(n)}$ (though, since we work with perturbed
surfaces in this paper, we will not deal with highly distorted cases
which might violate $\Theta_{(n)}<0$).  $\mathcal{S}$ will be assumed
to have spherical topology.  The time evolution of $\mathcal{S}$ has
been shown to be well behaved (locally in time) provided it satisfies
a suitable stability condition
\cite{Andersson:2007fh,Andersson:2005gq}; the 3-dimensional
hypersurface generated by this time evolution is thus smooth.
Isolated, dynamical and trapping horizons are all special cases of
such 3-dimensional hypersurfaces.  We shall not go into the detailed
definitions of these notions because we shall work only with
marginally outer trapped surfaces $\mathcal{S}$ at a single instant of
time.

We shall denote the spacelike 2-metric on $\mathcal{S}$ by $\q_{ab}$,
the covariant derivative compatible with $\q_{ab}$ is $\D_a$, the
intrinsic scalar curvature of $\q_{ab}$ is $\R$, and the invariant
volume 2-form is $\twoeps_{ab}$.  We shall work with $\mathcal{S}$ embedded
in a spatial hypersurface $\Sigma$.  The outgoing unit spacelike
normal to $\mathcal{S}$ within $\Sigma$ is denoted by $r^a$ and the
extrinsic curvature of $\Sigma$ embedded within the spacetime manifold
$\mathcal{M}$ is $K_{ab}$.  Another important field is the 1-form
$\tilde{\omega}_a = K_{cb}\q^c_ar^b$.  We assume that $\mathcal{S}$ is
axisymmetric, i.e. it admits a rotational symmetry $\varphi^a$ which
preserves $\q_{ab}$ and $\tilde{\omega}_a$
\cite{Dreyer:2002mx,Beetle:2008yt,Beetle:2013zga,Cook:2007wr,Lovelace:2008tw}.
Let $A_{\St}$ be the area of $\St$ and $R_\mathcal{S} =
\sqrt{A_{\St}/4\pi}$ its area radius.  The angular momentum associated
with $\mathcal{S}$ in vacuum general relativity is given by
\begin{equation}
  \label{eq:angmom}
  J_{\St}^{(\varphi)}= \frac{1}{8\pi}\oint_S \tilde{\omega}_a\varphi^a\,d^2V  = \frac{1}{8\pi}\oint_S K_{ab}\varphi^ar^b\,d^2V \,,
\end{equation}
where $d^2V$ is the invariant volume element on $\mathcal{S}$.  We
shall usually drop the superscript in $J_\mathcal{S}^{(\varphi)}$.
The mass associated with $\mathcal{S}$ is
\begin{equation}
  \label{eq:10}
  M_{\St} = \frac{1}{2R_\mathcal{S}}\sqrt{R_\mathcal{S}^4 + 4J_\mathcal{S}^2}\,.
\end{equation}
We shall need higher order multipoles beyond the mass and angular
momentum.  Multipole moments for isolated horizons were introduced in
\cite{Ashtekar:2004gp}.  A general procedure valid for dynamical black
holes without assuming symmetries is given in \cite{Ashtekar:2013qta}.
However, this requires access to the time evolution of $\mathcal{S}$
which is beyond the scope of this paper.  We shall therefore use a
simpler and more limited method described in \cite{Schnetter:2006yt}
which is a simple extension of \cite{Ashtekar:2004gp}.

The starting point for this method is to construct a preferred
coordinate system on $\mathcal{S}$ adapted to the axial symmetry:
$(\zeta,\phi)$, with $-1\leq\zeta\leq 1$ and $0\leq \phi<2\pi$.  We
normalize $\varphi^a$ so that it has affine length $2\pi$.  Then
$\phi$ is the affine parameter along $\varphi^a$: $\varphi^a\D_a\phi =
1$. The other coordinate $\zeta$ is defined by:
\begin{equation}
  \label{eq:11}
  \D_a\zeta = \frac{1}{R_\mathcal{S}^2}\varphi^b\twoeps_{ba}\,,\qquad \oint_{\mathcal{S}}\zeta\twoeps = 0\,.
\end{equation}
It can then be shown that in these coordinates the metric $\q_{ab}$
takes the form \cite{Ashtekar:2004gp}:
\begin{equation}
  \label{eq:12}
  \q_{ab} = R_\mathcal{S}^2\left( f^{-1}\D_a\zeta \D_b\zeta + f\D_a\phi \D_b\phi \right)\,,
\end{equation}
where $f$ is a function of $\zeta$: $f =
\varphi_a\varphi^a/R_\mathcal{S}^2$.  On a round sphere in Euclidean
space with the usual spherical coordinates $(\theta,\phi)$, $\zeta =
\cos\theta$.  Regularity of $\q_{ab}$ at the poles requires
\begin{equation}
  \label{eq:13}
  \lim_{\zeta\rightarrow \pm 1}f^\prime(\zeta) = \mp 2\,.
\end{equation}
It can also be shown that the scalar curvature is 
\begin{equation}
  \label{eq:14}
  \R = -\frac{1}{R_\mathcal{S}^2}f^{\prime\prime}(\zeta)\,.
\end{equation}
In these coordinates, the invariant volume element on $\mathcal{S}$ is
independent of $f$, and is thus the same as on a round 2-sphere where
$f = \sin^2\theta = 1-\zeta^2$.  The normalization condition for
spherical harmonics therefore works with the invariant volume element.
The mass multipoles are:
\begin{equation}
  \label{eq:15}
  \Mn = \frac{M_{\St} R_{\St}^n}{8\pi} \oint_\mathcal{S}P_n(\zeta)\R\,d^2V\,,
\end{equation}
and the angular momentum multipoles are
\begin{equation}
  \label{eq:16}
  \Jn = \frac{R_\mathcal{S}^{n+1}}{8\pi}\oint_\mathcal{S} \twoeps^{ab}\D_aP_n(\zeta)\, K_{bc}r^c\,d^2V\,.
\end{equation}
Here, $P_n(\zeta)$ are the Legendre polynomials.  On general
grounds, it follows that $\M{0} = M_{\St}$ and $\J{0} = 0$.  The
first angular momentum multipole is just the angular momentum: $\J{1} =
J_\mathcal{S}$. Furthermore, since $P_n(-\zeta) = (-1)^nP_n(\zeta)$,
if the horizon is axisymmetric and also symmetric under a reflection
($\zeta \rightarrow -\zeta$, $\phi\rightarrow \phi+\pi$), then $\Mn=0$
for all odd $n$ and $\Jn=0$ for all even $n$.

\subsection{The Kerr multipole moments}
\label{subsec:multipoles:kerr}

It is useful to illustrate these notions for a Kerr horizon parameterized
by a mass $m$ and spin parameter $a$.  We shall later compare the Kerr
multipole moments with the corresponding moments for a single spinning
Bowen-York black hole.

We first note that the horizon mass and spin are respectively $m$ and
$am$ as expected.  In Boyer-Lindquist coordinates $(t,r,\theta,\phi)$,
the horizon is located at $r=r_+$ such that (see
e.g. \cite{Chandrasekhar:1985kt})
\begin{equation}
  \label{eq:47}
  r_+ = m + \sqrt{m^2-a^2} \,.  
\end{equation}
The 2-metric on a cross-section of the horizon is
\begin{equation}
  \label{eq:48}
  \q_{ab} = \rho_+^2\nabla_a\theta\nabla_b\theta + \frac{(r_+^2+a^2)^2}{\rho_+^2}\sin^2\theta\nabla_a\phi\nabla_b\phi\,,
\end{equation}
where $\rho_+^2 = r_+^2+a^2\cos^2\theta$.  The volume 2-form on $S$ is
\begin{equation}
  \label{eq:49}
  \twoeps = (r_+^2+a^2)\sin\theta d\theta \wedge d\phi \,.
\end{equation}
The area of any closed cross-section of the horizon is $A =
4\pi(r_+^2+a^2)$ and the area radius is $R=\sqrt{r_+^2+a^2}$.  The
invariant coordinate $\zeta$ is, as for round 2-spheres, $\zeta =
\cos\theta$. The 2-metric can be written in the form given in
Eq.~(\ref{eq:12}) with
\begin{equation}
  \label{eq:50}
  f(\zeta) = \frac{R^2\sin^2\theta}{\rho_+^2} = \frac{1-\zeta^2}{1-(a/R)^2(1-\zeta^2)}\,.  
\end{equation}
It is easy to see that Eq.~(\ref{eq:13}) is satisfied. The general
calculation of the Kerr multipole moments is discussed in
\cite{Ashtekar:2004gp} and here we shall need the multipoles in the
limit of small spins.  Keeping terms up to $\mathcal{O}(a^2)$: 
\begin{equation}
  \label{eq:54}
  f(\zeta) = 1-\zeta^2 + \left(\frac{a}{R}\right)^2(1-\zeta^2)^2  + \mathcal{O}(a^4)\,,
\end{equation}
which implies
\begin{equation}
  \label{eq:55}
  \widetilde{R}(\zeta) = -\frac{1}{R^2}f^{\prime\prime}(\zeta) = \frac{2}{R^2} -\frac{8a^2}{R^4}P_2(\zeta) + \mathcal{O}(a^4)\,.
\end{equation}
Apart from the mass and angular momentum, the only non-vanishing
multipole moment at $\mathcal{O}(a^2)$ is
\begin{equation}
  \label{eq:56}
  \M{2} = -\frac{4}{5}ma^2 = -\frac{4J^2}{5m}\,.
\end{equation}
It is easier to calculate the $\Jn$ from the Weyl tensor component
$\Psi_2$ (in Boyer-Lindquist coordinates):
\begin{equation}
  \label{eq:57}
  \Psi_2 = -\frac{m}{(r-ia\cos\theta)^3}  = -\frac{m}{r^3}\left(1 + \frac{3ia}{r}\cos\theta + \frac{6a^2}{r^2}\cos^2\theta + \mathcal{O}(a^3)\right)\,.
\end{equation}
For isolated horizons in vacuum general relativity, it can be shown
that $\R = -4\mathrm{Re}[\Psi_2]$
\cite{Ashtekar:1999yj,Ashtekar:2000hw}.  Writing $m$ and $r$ in terms
of $R$ and $a$, Eq.~(\ref{eq:55}) is recovered.  The moments of the
imaginary part of $\Psi_2$ yield the angular momentum multipole
moments $\Jn$ \cite{Ashtekar:2004gp}.  It is again easy to see that
$\J{0} =0$ and $\J{1}=J=am$.  All other moments vanish at this order
of approximation.

\subsection{Perturbations of the multipole moments}
\label{subsec:multipoles:pert}

Consider a black hole with mass $M_1$, angular momentum $J_1$, and
multipole moments $\Mn$, $\Jn$.  We shall be concerned with how the
multipole moments change under the influence of an external
perturbation. Let $\delta \Mn$ and $\delta \Jn$ be the changes in
$\Mn$ and $\Jn$ respectively. Consider an external perturbation caused
by a non-spinning binary companion of mass $M_2$ placed at a distance
$d$.  We shall restrict ourselves to axisymmetric situations where the
separation vector between the two black holes is parallel to the
spin-vector of the first black hole.  Let us define the dimensionless
spin of the first black hole as $\chi = J_1/M_1^2$; $\chi$ can be
shown to be restricted to $|\chi| < 1$ \cite{Jaramillo:2011pg}. We
assume that the full set of multipole moments is fully determined by
the lowest non-vanishing moments $M_1$ and $J_1$, i.e. the mass and
the angular momentum.

The small quantities in the problem are $M_2$, $1/d$ and $\chi$, and
we shall consider dimensionless combinations of these parameters.  On
general grounds, $\delta \Mn$ and $\delta \Jn$ can be expanded as
\begin{eqnarray}
  \label{eq:25}
  \frac{\delta \Mn}{M_{1}^{n+1}} = \sum_{m,k,j =1}^\infty\alpha^{(n)}_{mkj} \frac{M_{1}^m M_{2}^k}{d^{m+k}} \chi^j\,,\\
  \frac{\delta \Jn}{M_{1}^{n+1}} = \sum_{m,k,j = 1}^\infty\beta^{(n)}_{mkj} \frac{M_{1}^m M_{2}^k}{d^{m+k}} \chi^j\,.
\end{eqnarray}
The dimensionless coefficients $\alpha^{(n)}$ and $\beta^{(n)}$ will
be called tidal coefficients.  The masses $M_{1,2}$ are the physical
masses and will be combinations of the ``bare'' parameters of the
system which might include the bare masses $m_{1,2}$, $d$ and the
angular momenta; we shall see explicit examples of this later.  

If $M_2\rightarrow 0$ or $d\rightarrow \infty$, then the external
perturbation vanishes and hence ($\delta \Mn$, $\delta \Jn$) must also
vanish.  This means that in the above sums, we need only consider
$k,m+k \geq 1$.  Similarly, we do not expect a divergence when
$\chi,M_1 \rightarrow 0$, which shows that $m,j\geq 1$.  Thus, all the
exponents $(m,k,j)$ can take only positive values.  We expect
additional terms if the second black hole were also spinning, and if
both black holes had non-zero linear momentum.  In non-axisymmetric
systems, when for example the angular and linear momenta, and the
separation vector, are not aligned, we would have to consider moments
($\M{\ell m},\J{\ell m}$) for $m\neq 0$ as well.  These
generalizations will be discussed in a forthcoming publication.

It is also useful to note that in the non-spinning case, the
perturbations start to be non-vanishing only from
$\mathcal{O}(M_2/d^3)$ onwards.  Thus, the first term in, say
$\M{2}$, is proportional to $M_1^2M_2/d^3$ and the corresponding tidal
coefficient is $\alpha^{(2)}_{210}$.  This coefficient will be
proportional to the tidal Love number $h_2$ calculated in
\cite{Damour:2009va,Landry:2014jka}.

\section{The Brill-Lindquist and Bowen-York initial data sets}
\label{sec:bowenyork}

We work in a 3+1 split of spacetime where initial data are specified on
a spacelike hypersurface $\Sigma$.  The initial data consist of the
positive-definite three metric $h_{ab}$ on $\Sigma$, and the extrinsic
curvature $K_{ab}$ describing the embedding of $\Sigma$ within the
spacetime manifold $\mathcal{M}$. The initial data
$(\Sigma,h_{ab},K_{ab})$ satisfy the momentum and Hamiltonian
constraint equations respectively:
\begin{equation}
  \label{eq:1}
  D_a(K^{ab}-Kh^{ab}) = 0\,,\qquad {}^3R - K^{ab}K_{ab} + K^2 =0\,.
\end{equation}
Here ${}^3R$ is the Ricci scalar computed from $h_{ab}$ and $D_a$ is
the covariant derivative compatible with $h_{ab}$.  

We shall take $h_{ab}$ to be conformally flat so that $h_{ab}=\psi^4
f_{ab}$, with $f_{ab}$ being a flat metric, and furthermore, we shall
take $K_{ab}$ to be trace-free: $K=0$. With these choices, the
constraint equations become
\begin{equation}
  \label{eq:4}
  \Delta\psi = -\frac{1}{8}\psi^{-7}\K_{ab}\K^{ab}\,,\qquad \partial_a\K^{ab} = 0\,.
\end{equation}
Here $\K_{ab} = \psi^2K_{ab}$ is the re-scaled extrinsic curvature,
$\Delta := \partial_a\partial^a$ is the flat-space Laplacian and
$\partial_a$ is the derivative operator compatible with the flat
metric $f_{ab}$.  Since the momentum constraint is now decoupled, we
use an appropriate solution $\K_{ab}$ to the momentum constraint, plug
it into the Hamiltonian constraint and solve the resulting elliptic
equation for $\psi$.  Furthermore, since the momentum constraint is
seen to be linear, we can linearly superpose various solutions.  The
Hamiltonian constraint however is non-linear and introduces various
cross-terms between the different pieces included in $\K_{ab}$.

The simplest solutions are when the extrinsic curvature vanishes
identically so that the data are time-symmetric. In this case the
conformal factor satisfies the Laplace equation in flat space.
Non-trivial solutions are obtained when we have ``point charges''.
Thus, if we place $N$ masses $m_i$ at points $\mathbf{r}_i$
respectively ($i=1\ldots N$), then at any position $\mathbf{r}$ on
$\Sigma$ away from $\mathbf{r}_i$:
\begin{equation}
  \label{eq:5}
  \psi_{BL}(\mathbf{r}) = 1 + \sum_{i=1}^N\frac{m_i}{2|\mathbf{r}-\mathbf{r}_i|}\,.
\end{equation}
This is the well known Brill-Lindquist solution \cite{Brill:1963yv};
see also \cite{Misner:1957mt,Gibbons:1972ym}. We shall consider the
cases of a single black hole or a binary system so that the sum over
$i$ is either just a single term or the sum of two.  The parameters
$m_i$ are the \emph{bare} masses of the black holes.  In the absence
of any other black hole, these would be the physical horizon mass
(and also the ADM mass).  However, this is not the case if other black
holes are present.

Linear momentum $\mathbf{P}$ and angular momentum $\mathbf{J}$ for a
single black hole are handled by non-trivial choices for $\K_{ab}$,
denoted by ${}^P\K_{ab}$ and ${}^J\K_{ab}$ \cite{Bowen:1980yu}:
\begin{eqnarray}
  \label{eq:6}
  {}^P\K_{ab} &=& \frac{3}{2r^2}\left[ P_an_b + P_bn_a - (f_{ab} - n_an_b)P_cn^c\right]\,,\\
  \label{eq:18}
  {}^J\K_{ab} &=& \frac{3}{r^3}\left[ \epsilon_{acd}J^cn^dn_b + \epsilon_{bcd}J^cn^dn_a\right] \,.
\end{eqnarray}
These are the well known Bowen-York solutions to the momentum
constraints.  Here we have chosen standard spherical coordinates
centered on the location of the black hole, with $r$ as the radial
coordinate and $n^a$ the unit 3-vector orthogonal to the spheres of
constant $r$.  Solutions with multiple black holes are obtained by
superposing the different individual extrinsic curvatures.  The
solution for the conformal factor is, however, a non-linear combination
of the extrinsic curvatures.

We study the effects of momentum, spin and presence of a binary
companion on a black hole by considering perturbative solutions to the
conformal factor \cite{Brandt:1997tf}:
\begin{equation}
  \label{eq:7}
 \psi = \psi_{BL} + u\,.
\end{equation}
This is the so-called puncture ansatz, where $\psi_{BL}$ contains all
the singularities in the conformal factor and $u$ is taken to be
smooth everywhere and vanishing at spatial infinity.  The equation for
the conformal factor becomes:
\begin{equation}
  \label{eq:8}
  \tilde{\Delta}u = -\frac{1}{8}(\psi_{BL}+ u)^{-7}\K_{ab}\K^{ab}\,.
\end{equation}
We shall keep terms up to $\mathcal{O}(P^2)$, $\mathcal{O}(J^2)$
and $\mathcal{O}(PJ)$.  It is easy to see that at this level of
accuracy (since $\K_{ab}\K^{ab}$ contains only terms of this order),
the conformal factor satisfies a linear Poisson equation:
\begin{equation}
  \label{eq:9}
  \tilde{\Delta}u = -\frac{1}{8}\psi_{BL}^{-7}\K_{ab}\K^{ab}\,.
\end{equation}
Even with this simplification, the right hand side of this equation is
fairly complicated and it contains various cross terms between the
spin and linear-momenta (of either black hole in the case of a binary
system).  Still, given its linearity, we can treat it analytically.
We would get a linear equation if we keep terms linear in $u$ on the
right-hand-side of Eq.~(\ref{eq:8}).  This case would still be
amenable to an analytic treatment and would allow us to go to higher
orders in $P$ and $J$, but we shall restrict ourselves to dropping all
$u$ dependence within the source term.

\section{A single spinning black hole}
\label{sec:singlebh}

\subsection{The conformal factor}

The solution to the momentum constraint for a single spinning black
hole at rest and placed at the origin is given by Eq.~(\ref{eq:18}).
A simple calculation shows
\begin{equation}
  \label{eq:17}
  \K_{ab}\K^{ab} = \frac{18J^2 \sin^2\theta}{r^6}\,.
\end{equation}
The angle $\theta$ is measured from $\mathbf{J}$.  Note again that the
parameters $m$ and $J$ are the bare mass and angular momentum
respectively.  The physical parameters (either at the horizon or at
spatial infinity) will be determined below.  With the puncture ansatz
of Eq.~(\ref{eq:7}), the Hamiltonian constraint becomes
\begin{eqnarray}
  \nabla^2u &=& -\frac{\psi^{-7}}{8}\K_{ab}\K^{ab} = -\frac{288rJ^2\sin^2\theta}{(m+2r + 2ur)^7} \nonumber \\ &\approx& -\frac{192J^2r}{(m+2r)^7}\left( 1-P_2(\cos\theta)\right)\,.
  \label{eq:19}
\end{eqnarray}
In the last step, as explained earlier, we have dropped the $u$
dependent terms on the right hand side, and the resulting Poisson
equation is thus only valid up to $\mathcal{O}(J^2)$.  We require that
$u$ is regular and $u\rightarrow 0$ when $r\rightarrow\infty$.  Since
we are working in spherical coordinates, regularity at the origin
implies
\begin{equation}
  \label{eq:20}
  \left.\frac{\partial u}{\partial r}\right|_{r=0} = 0\,.
\end{equation}
The solution $u(r,\theta)$ will be of the form 
\begin{equation}
  \label{eq:21}
  u(r,\theta) = u_0(r)P_0(\cos\theta) + u_2(r)P_2(\cos\theta)\,,
\end{equation}
and the radial equations for $u_0(r)$ and $u_2(r)$ are: 
\begin{eqnarray}
  \label{eq:22}
  && u_0^{\prime\prime} + \frac{2}{r}u_0^\prime = -\frac{192J^2r}{(m+2r)^7}\,,\\
  && u_2^{\prime\prime} + \frac{2}{r}u_2^\prime -\frac{6}{r^2}u_2 = \frac{192J^2r}{(m+2r)^7}\,.
\end{eqnarray}
The solutions which are regular at $r=0$ and asymptotically flat are: 
\begin{eqnarray}
  \label{eq:26}
  u_0(r) &=& \frac{2J^2}{5m^3(m+2r)^5}(m^4 + 10m^3r + 40m^2r^2 + 40mr^3 + 16r^4)\,,\\
  \label{eq:71}
  u_2(r) &=& -\frac{16J^2r^2}{5m(m+2r)^5}\,.
\end{eqnarray}
We see that for large $r$, $u_2$ falls off as $1/r^3$ and
\begin{equation}
  \label{eq:27}
  u_0(r) = \frac{2J^2}{5m^3}\frac{1}{2r} + \mathcal{O}(1/r^2)\,.
\end{equation}
Thus, the ADM mass is, ignoring higher powers in $J$,
\begin{equation}
  \label{eq:28}
  m_{ADM} = m + \frac{2J^2}{5m^3} \,.
\end{equation}
The values of $u_0$ and $u_2$ at $r=m/2$ will be used later. These are: 
\begin{equation}
  \label{eq:67}
  \left. u_0\right|_{r=m/2} = \frac{11}{40} \frac{J^2}{m^4}\,,\qquad \textrm{and}\qquad \left. u_2\right|_{r=m/2} = -\frac{1 }{40}\frac{J^2}{m^4}\,. 
\end{equation}

\subsection{The location of the marginal surface}

Let us now turn to the location of the marginal surface $\St$.  We
need to find closed surface(s) within $\Sigma$ such that the outward
null normal has vanishing expansion.  If $r^a$ is the outward
spacelike unit-normal to $\St$ within $\Sigma$, and $\tau^a$ is the
unit timelike normal to $\Sigma$, then all outward null normals are
parallel to $\ell^a = \tau^a + r^a$. Thus, $\St$ is a marginally outer
trapped surface if
\begin{equation}
  \label{eq:2}
  q^{ab}\nabla_a\ell_b = (h^{ab}-r^ar^b)\nabla_a(\tau_b+r_b) = D_ar^a + K - K_{ab}r^ar^b = 0\,.  
\end{equation}
The general solution is given by $f(r,\theta) = r-h(\theta) = 0$.
Cook and York have previously studied the horizon location for a
single spinning and boosted Bowen-York black hole \cite{Cook:1989fb}.
It is however useful for us to repeat some of the calculations for the
zero-boost case.

The unit normal to $S$ is
\begin{equation}
  \label{eq:23}
  \mathbf{r} = \frac{\psi^{-2}}{\sqrt{1+h_\theta^2/r^2}}\left( \partial_r - \frac{h_\theta}{r^2}\partial_\theta \right)
\end{equation}
It is easy to check that $K_{ab}r^ar^b = 0$.  The horizon is thus a
minimal surface and, more explicitly, it is obtained by solving
\begin{equation}
  \label{eq:24}
  \frac{\partial}{\partial r}\left( \frac{r^2\psi^4}{\sqrt{1+h_\theta^2/r^2}}\right) = \frac{1}{\sin\theta}\frac{\partial}{\partial\theta}\left( \frac{\psi^4\sin\theta h_\theta}{\sqrt{1+h_\theta^2/r^2}} \right) \,.
\end{equation}
We can solve this order by order in $J$.  If all terms in $J$ are
ignored, Eq.~(\ref{eq:24}) becomes $\partial_r(r^2\psi_{BL}^4) =0$
whose solution is $r=m/2$ \cite{Brill:1963yv}.  Now keep terms linear
in $h_\theta$ (this can be, at best, linear in $J$), and dropping all
terms beyond $\mathcal{O}(J^2)$, Eq.~(\ref{eq:24}) becomes
\begin{equation}
  \label{eq:62}
  \left.\frac{d}{dr}(r^2\psi_{BL}^4)\right|_{r=h(\theta)} +  \left.\frac{d}{dr}(4r^2\psi_{BL}^3u)\right|_{r=h(\theta)} = \frac{16}{\sin\theta}\frac{d}{d\theta}\left(\sin\theta\frac{d h}{d\theta} \right)\,.
\end{equation}
In the second term on the left, the derivative can be evaluated at
$r=m/2$ since $u$ is already $\mathcal{O}(J^2)$.  Using the solutions
for $u_0$ and $u_2$ derived earlier, it turns out somewhat
surprisingly that this term vanishes.  As for the first term:
\begin{eqnarray}
  \label{eq:63}
  \left.\frac{d}{dr}(r^2\psi_{BL})\right|_{r=h(\theta)} &=& \left.\frac{d}{dr}(r^2\psi_{BL})\right|_{r=m/2} + \left(h-\frac{m}{2}\right)\left.\frac{d^2}{dr^2}(r^2\psi_{BL})\right|_{r=m/2} 
  \nonumber
  \\ &=& 16\left(h-\frac{m}{2}\right)\,. 
\end{eqnarray}
Putting it all together, it is easily seen that the solution to
Eq.~(\ref{eq:62}) is just $h=m/2$. A similar calculation shows that
this holds also at $\mathcal{O}(J^2)$.

\subsection{The area, angular momentum, and mass}

With the location of the marginal surface $\St$ known, we can turn to
its physical and geometrical properties.  The first is simply its
area.  The induced metric on a surface given by $r=h(\theta)$
is
\begin{equation}
  \label{eq:64}
  ds^2_{\q} = \psi^4\left((r^2 + h_\theta^2)d\theta^2 + r^2\sin^2\theta d\phi^2\right)\,.
\end{equation}
The invariant volume measure is 
\begin{equation}
  \label{eq:59}
  \sqrt{\textrm{det}\q} =
r^2\psi^4\sin\theta\sqrt{1+h_\theta^2/r^2}\,.
\end{equation}
Specializing to the marginal surface $r=m/2$ found earlier, and keeping
terms up to $\mathcal{O}(J^2)$, we get
\begin{equation}
  \label{eq:65}
  \sqrt{\textrm{det}\q} \approx \frac{m^2}{4}(\left.\psi_{BL} + u)^4\right|_{r=m/2}\sin\theta \approx 4m^2(1+2\left.u\right|_{r=m/2})\sin\theta\,.
\end{equation}
The area is thus 
\begin{eqnarray}
  \label{eq:3}
  A &=& 2\pi \int_0^\pi \sqrt{\textrm{det}\q}\,d\theta  \approx 8\pi m^2 \int_0^\pi (1+2\left.u\right|_{r=m/2})\sin\theta\,d\theta \\ 
  &=& 16\pi m^2 \left(1+2\left.u_0\right|_{r=m/2}\right) = 16\pi m^2 \left(1+\frac{11J^2}{20m^4}\right) \,,
\end{eqnarray}
and the corresponding area radius is 
\begin{equation}
  \label{eq:66}
  R = \sqrt{\frac{A}{4\pi}} \approx 2m\left(1+\frac{11J^2}{40m^4}\right)\,.
\end{equation}
The angular momentum turns out to be just the parameter $J$ appearing
in the extrinsic curvature.  To see this, consider any surface
$r=h(\theta)$ ($h(\theta)$ could be arbitrary, subject only to the
condition that the surface is smooth and of spherical topology).
Then, taking all the factors of $\psi$ into account, we get
\begin{equation}
  \label{eq:58}
  K_{ab}r^a\varphi^b = -\frac{3\psi^{-4}J\sin^2\theta}{r^2\sqrt{1+h_\theta^2/r^2}}\,.
\end{equation}
Using Eq.~(\ref{eq:59}), it can be shown that the angular momentum
associated with the marginal surface (given by Eq.~(\ref{eq:angmom}))
is just $J$.  Similarly, this shows that the ADM angular momentum
associated with the sphere at spatial infinity is also $J$.  This fact
can also be seen by the balance law for angular momentum discussed in
\cite{Ashtekar:2004cn}, obtained by integrating the momentum constraint
over $\Sigma$ after a contraction with $\varphi^a$.  Using the fact
that $\varphi^a$ is a symmetry of $h_{ab}$ then shows that the angular
momentum for any closed spherical 2-surface is $J$.

Using Eqs.~(\ref{eq:10}) and (\ref{eq:66}), the mass of the horizon is
\begin{equation}
  \label{eq:29}
  M \approx \frac{R}{2}\left(1+\frac{2J^2}{R^4}\right) =  m\left(1+\frac{2J^2}{5m^4}\right) + \mathcal{O}(J^4)\,.
\end{equation}
We have dropped the subscript $\St$ on $M$ for simplicity.
Henceforth, we shall usually use $M$ for the horizon mass to
distinguish it from the bare mass $m$.  It is interesting to note that
the value obtained here is the same as the ADM mass given in
Eq.~(\ref{eq:28}).  We are now ready to turn to the higher multipole
moments.

\subsection{Higher multipole moments}
\label{subsec:singhebh:moments}

In order to calculate the multipole moments $(\Mn,\Jn)$ we first need
to find the preferred coordinate system $(\zeta,\phi)$ compatible with
the axial symmetry.  Starting with Eq.~(\ref{eq:64}), keeping terms up
to $\mathcal{O}(J^2)$, we see that the metric $ds^2_\q$ can be put in
the form of Eq.~(\ref{eq:12}) with 
\begin{equation}
  \label{eq:68}
  f = \frac{r^2\psi^4}{R^2}\sin^2\theta \,\qquad \textrm{and} \qquad d\zeta = \frac{r^2\psi^4}{R^2}\sin\theta\,d\theta\,.
\end{equation}
It is useful to again note that at $r=m/2$, $r^2\psi^4 \approx
4m^2(1+2u)$.  Setting $r=m/2$, using the values of $u_0$ and $u_2$ at
$r=m/2$, and the result for $R$, it is not difficult to show that
\begin{equation}
  \label{eq:30}
  \zeta = \cos\theta \left(1+\frac{J^2\sin^2\theta}{40m^4} \right) + \mathcal{O}(J^4)\,.
\end{equation}
As expected, $\zeta = 1$ and $-1$ at the north and south poles
respectively. It is also easy to check that the condition of
Eq.~(\ref{eq:13}) is indeed satisfied.  We use Eq.~(\ref{eq:14}) to
calculate the scalar curvature $\R$.  We begin with:
\begin{equation}
  \label{eq:69}
  \frac{df}{d\zeta} = \frac{df}{d\theta}\frac{d\theta}{d\zeta} = \frac{(1+2u)2\sin\theta\cos\theta + 2u_\theta\sin^2\theta}{(1+2u)\sin\theta} \approx 2\cos\theta + 2u_\theta\sin\theta\,.
\end{equation}
A similar further short calculation, utilizing also Eq.~(\ref{eq:67}),
yields the intrinsic scalar curvature of the horizon:
\begin{equation}
  \label{eq:31}
  \R = -\frac{1}{R^2}\frac{d^2f}{d\zeta^2} = \frac{2}{R^2} - \frac{J^2}{20M^6}P_2(\zeta) + \mathcal{O}(J^4)\,.  
\end{equation}
This finally allows us to calculate the mass multipole moments. At the
approximation that we are working in, the mass quadrupole moment is
\begin{equation}
  \label{eq:32}
  \M{2} = -\frac{2J^2}{25M} + \mathcal{O}(J^4)\,.  
\end{equation}
All higher moments $\Mn$ vanish.  Similarly, it turns out that the
only non-vanishing angular momentum multipole $\Jn$ within our
approximation is the angular momentum $\J{1}$. All other $\Jn$ vanish
up to $\mathcal{O}(J^2)$.

It is interesting to compare these results with the corresponding
moments for the Kerr black hole horizon. The only one we can compare
is the mass-quadrupole $\M{2}$.  Comparing Eqs.~(\ref{eq:32}) and
(\ref{eq:56}), we see that the Kerr value is exactly 10 times larger;
thus the Bowen-York black hole is in fact closer to the Schwarzschild
black hole (with the same mass).

\subsection{A single boosted and spinning black hole}
\label{subsec:singlebh:boost}

Let us now consider a single Bowen-York black hole with non-vanishing
boost, i.e. including the solution ${}^P\K_{ab}$ of the momentum
constraint given in Eq.~(\ref{eq:6}).  For the moment, let us set the
angular momentum to zero and consider a non-spinning boosted black
hole.  Let us align the z-axis with the linear momentum $\mathbf{P}$
and, for an arbitrary point $P$ away from the puncture $r=0$, let
$\theta$ be the angle between the position vector $\mathbf{r}$ of $P$
and the z-axis. Then, it is easy to show that
\begin{equation}
  \label{eq:41}
  \K_{ab}\K^{ab} = \frac{9P^2}{r^4}\left(\frac{1}{2} + \cos^2\theta\right)\,.  
\end{equation}
A perturbative solution to the Hamiltonian constraint for small $P$
has been obtained previously \cite{Cook:1989fb,Dennison:2006nq}.  The
calculations are very similar to what we have seen for the spinning
case earlier and we shall not repeat all the steps here.  As shown in
\cite{Dennison:2006nq}, with the puncture ansatz, the correction term
for the conformal factor is
\begin{equation}
  \label{eq:42}
  u(r,\theta) = \epsilon_P^2 \left( u_0(r)P_0(\cos\theta) + u_2(r)P_2(\cos\theta) \right) + \mathcal{O}(\epsilon_P^4)\,. 
\end{equation}
Here $\epsilon_P := P/m$.  The solutions for the radial functions
$u_0$ and $u_2$ which are regular everywhere and asymptotically flat
are given in Eqs.~(A8) and (A9) of \cite{Dennison:2006nq}.  The
marginally trapped surface is located in Eq.~(24) of
\cite{Dennison:2006nq}:
\begin{equation}
  \label{eq:43}
  r = h(\theta) = \frac{m}{2} - \frac{P}{16}\cos\theta + \mathcal{O}(\epsilon_P^2)\,.
\end{equation}
We can easily calculate the multipole moments of this horizon.  The
horizon mass, in this case this is just the irreducible mass, has
already been calculated in \cite{Dennison:2006nq}:
\begin{equation}
  \label{eq:45}
  M = m\left(1+\frac{P^2}{8m^2}\right) + \mathcal{O}(P^4)\,.
\end{equation}
The angular momentum multipoles all vanish, and the mass quadrupole
moment turns out to be:
\begin{equation}
  \label{eq:46}
  \M{2} = \frac{MP^2}{200}(1871-2688\ln[2]) + \mathcal{O}(P^4)\,.
\end{equation}
Let us now combine these results with the results of the previous
sections on the single spinning black hole.  We shall restrict
ourselves to the axisymmetric situation with $\mathbf{P}$ and
$\mathbf{J}$ parallel to each other.  Then, from the form of the spin
and momentum contributions to the extrinsic curvature, a short
calculation shows that
\begin{equation}
  \label{eq:44}
  \K_{ab}\K^{ab} = \frac{9P^2}{r^4}\left(\frac{1}{2}+\cos^2\theta\right) + \frac{18}{r^6}J^2\sin^2\theta\,.
\end{equation}
There are no cross terms between the angular and linear momenta; this
would cease to hold if $\mathbf{P}$ and $\mathbf{J}$ were not aligned.
We again look for solutions of the form given in Eq.~(\ref{eq:21}).
Solutions to the two radial equations which are regular and
asymptotically flat are obtained by linearly superposition of
Eqs.~(\ref{eq:26}-\ref{eq:27}) with the corresponding solutions given
in \cite{Dennison:2006nq}. The location of the horizon is still given
by Eq.~(\ref{eq:43}). The only non-vanishing multipole moment apart
from the mass and the spin (at the approximation we are working in) is
the mass quadrupole moment, which is the sum of the pure spin and boost
values given in Eqs.~(\ref{eq:32}) and (\ref{eq:46}).

\section{A spinning black hole with a non-spinning binary companion}
\label{sec:binary}

We now place our spinning black hole in a binary system.  We shall
simplify our calculation in three ways. First, we shall ignore the
effects of linear momentum.  Second, we shall take the companion black
hole to be non-spinning and finally, we shall take the separation
vector between the two black holes to be parallel to the angular
momentum vector $\mathbf{J}$.  With these restrictions, the initial
data is guaranteed to be axisymmetric.  While not trivial, it is in
fact not hard to relax these assumptions since we have a flat
background metric available to us.  However, breaking axial symmetry
introduces complications in the definitions of the multipole moments
and calls for a separate discussion.  We shall address this in a
forthcoming paper.  Moreover, as in the earlier sections, we shall
work in the limit of small angular momentum (including terms accurate
to $\mathcal{O}(J^2)$); this restriction is however difficult to avoid
in an analytic treatment and numerical calculations will be required
for more accuracy.
\begin{figure}
  \centering
  \includegraphics[width=0.6\textwidth]{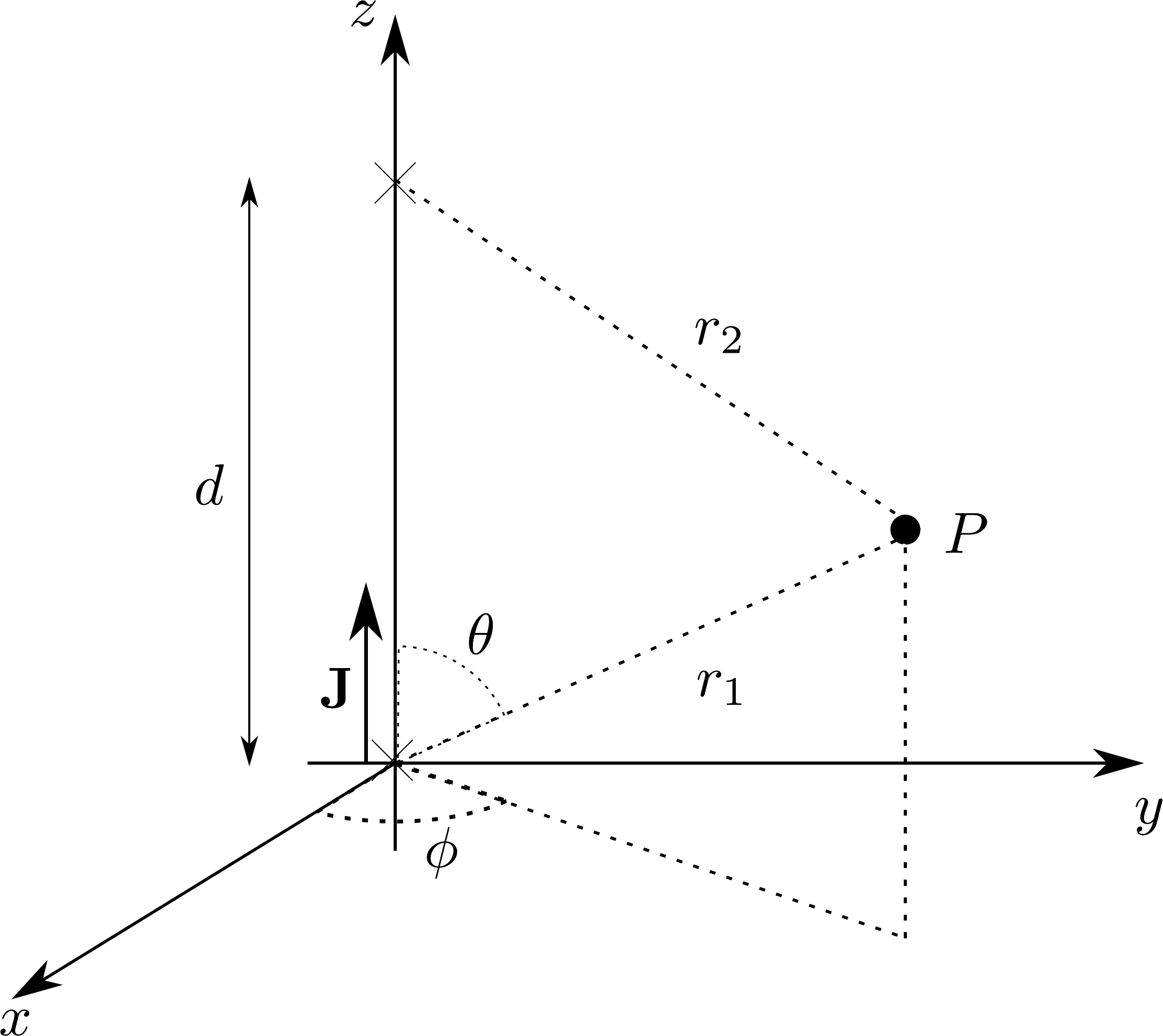}
  \caption{A depiction of the binary system.  The first black hole is
    placed at the origin and it's angular momentum $\mathbf{J}$ is
    aligned with the $z$-axis.  The second black hole is placed at a
    distance $d$ on the $z$-axis. The distances of an arbitrary point
    $P$ from the two black holes are $r_1$ and $r_2$, and the angular
    coordinates $(\theta,\phi)$ of $P$ are defined in the usual way.}
  \label{fig:binary}
\end{figure}

When the spins vanish identically and we have time symmetry, the exact
solution to the Hamiltonian constraint is given by the Brill-Lindquist
solution (Eq.~(\ref{eq:5})):
\begin{equation}
  \label{eq:33}
  \psi_{BL}(\mathbf{r}) := \frac{1}{\alpha} = 1 + \frac{m_1}{2r_1} + \frac{m_2}{2r_2}\,.
\end{equation}
With the puncture ansatz of Eq.~(\ref{eq:7}), again ignoring $u$ in
the source term, we obtain:
\begin{equation}
  \label{eq:34}
    \tilde{\Delta}u = -\frac{1}{8}\alpha^7\K_{ab}\K^{ab}\,.
\end{equation}
We take the second black hole to be non-spinning so that it does not
have any contribution to the extrinsic curvature and all the
dependence on $m_2$ and $d$ is through $\alpha$ in the above equation.
From the Bowen-York extrinsic curvature we get explicitly:
\begin{equation}
  \label{eq:37}
  \tilde{\Delta}u = -\frac{9}{4r^6}\alpha^7J^2\sin^2\theta + \mathcal{O}(J^4)\,.
\end{equation}
We could, in principle, choose to keep terms up to any order in $1/d$
that we wish.  Since tidal effects (in the absence of spin) are
proportional to $m_2/d^3$, we shall keep our calculations accurate to
$\mathcal{O}(1/d^3)$.  We start by expanding $\alpha$ in terms of
Legendre polynomials:
\begin{eqnarray}
  \label{eq:35}
  \frac{1}{\alpha} &=& 1 + \frac{m_1}{2r} + \frac{m_2}{2\sqrt{r^2+d^2-2dr\cos\theta}} \nonumber \\
  &=& 1+\frac{m_1}{2r} + \frac{m_2}{2d}\sum_{n=0}^\infty P_n(\cos\theta)\left(\frac{r}{d}\right)^n\,.
\end{eqnarray}
We have chosen to expand in powers of $1/d$ because we are interested
in the region near the first black hole, i.e. near the origin where
$r$ can be small.  We should not expect the solutions we obtain using
this approximation to be uniformly accurate for large $r$.

If we keep terms up to $\mathcal{O}(1/d^3)$, we see that $\alpha$ will
include Legendre polynomials up to $P_2(\cos\theta)$.  Since
$\alpha^7$ is multiplied by $\sin^2\theta$ in Eq.~(\ref{eq:37}), it is
clear that the source term in that equation will include terms up to
$P_4(\cos\theta)$.  We thus look for a solution of the form
\begin{equation}
  \label{eq:38}
  u(r,\theta) = \sum_{n=0}^4 u_n(r)P_n(\cos\theta)\,.  
\end{equation}
Substituting this in Eq.~(\ref{eq:37}) then leads to five linear ODEs
for each of the five radial functions $u_n(r)$.  We display explicitly
the five differential equations (each equation is accurate up to
correction terms, which are $\mathcal{O}(J^4)$ and
$\mathcal{O}(1/d^4)$, and we define $\beta := m_1+2r$):
\begin{eqnarray}
  \label{eq:36}
  \frac{d}{dr}\left( r^2\frac{du_0}{dr}\right) &=& -\frac{192J^2r^3}{\beta^7}\left[ 1 -\frac{7m_2r}{\beta d} + \frac{28m_2^2r^2}{\beta^2d^2} + \frac{7m_2r^3}{5\beta^3d^3}(\beta^2-60m_2^2)\right]\,,\\ 
  \label{eq:36a}
  \frac{d}{dr}\left( r^2\frac{du_1}{dr}\right) -2u_1 &=& \frac{4032J^2m_2r^5}{5\beta^8}\left[ \frac{1}{d^2} - \frac{8m_2r}{\beta d^3} \right]\,,\\
  \label{eq:36b}
  \frac{d}{dr}\left( r^2\frac{du_2}{dr}\right) -6u_2 &=& \frac{192J^2r^3}{\beta^7}\left[ 1 - \frac{7m_2r}{\beta d} + \frac{28m_2^2r^2}{\beta^2 d^2}  \right. \nonumber \\ 
  && \left. + \frac{m_2r^3}{\beta^3d^3}(5\beta^2-84m_2^2)\right] \,,\\
  \label{eq:36c}
  \frac{d}{dr}\left( r^2\frac{du_3}{dr}\right) -12u_3 &=& -\frac{4032J^2m_2r^5}{5\beta^8}\left[ \frac{1}{d^2} - \frac{8m_2r}{\beta d^3}\right] \,, \\
  \label{eq:36d}
  \frac{d}{dr}\left( r^2\frac{du_4}{dr}\right) -20u_4 &=& -\frac{3456J^2m_2r^6}{5\beta^8d^3}\,.
\end{eqnarray}
The solutions which are regular at the origin are given in \ref{sec:appendix}.

\subsection{The marginal surface and multipole moments}
\label{subsec:binary:moments}

The procedure for calculating the multipole moments is the same as
before.  First we locate the marginal surface, find the
axially-symmetric geometry (i.e. the coordinate $\zeta$), expand the
scalar curvature and $\widetilde{\omega}_a$ in terms of the Legendre
polynomials and read off the multipole moments.  Since the methods
employed for each of these steps are technically very similar to what
was done in the previous section, we shall skip most of the
intermediate details and mostly provide results.  We shall start with
the non-spinning case and include spin effects subsequently.

\subsubsection{Non-spinning black holes}

For two non-spinning black holes, the conformal factor is known
exactly and is just the Brill-Lindquist result $\Psi_{BL}$ given in
Eq.~(\ref{eq:33}).  As before, we expand this in powers of $1/d$ given
in Eq.~(\ref{eq:35}) and keep terms up to $\mathcal{O}(1/d^3)$.  The
marginal surface is again found by solving Eq.~(\ref{eq:24}).  This
time we proceed order-by-order in $1/d$; details are provided in
Appendix B of \cite{BK-thesis}.  The location of the horizon is: 
\begin{eqnarray}
  \label{eq:51}
  r &=& h(\theta) = \frac{m_1}{2}\left[ 1 - \frac{m_2}{2d} + \frac{m_2}{4d^2}\left(m_2-m_1P_1(\cos\theta)\right)\right. \nonumber \\
  && \left. - \frac{m_2}{8d^3}\left(m_2^2 - 3m_1m_2P_1(\cos\theta) + \frac{5}{7}m_1^2P_2(\cos\theta)\right)\right] + \mathcal{O}(d^{-4})\,.
\end{eqnarray}
The angular dependence starts only from $1/d^2$ onwards.  In order to
find the area of the marginal surface and its geometric properties, we
need to evaluate $r^2\psi^4$ at the horizon accurate to
$\mathcal{O}(1/d^3)$:
\begin{equation}
  \label{eq:52}
  \left(r^2\psi^4\right)_{r=h(\theta)} \approx 4m_1^2 \left[ \left( 1+\frac{m_2}{2d}\right)^2 + \frac{m_1m_2}{2d^2}P_1(\cos\theta) + \frac{m_1^2m_2}{4d^3}P_2(\cos\theta) \right]\,.
\end{equation}
The area is then
\begin{equation}
  \label{eq:53}
  A = \int_0^\pi \sin\theta\,d\theta \int_0^{2\pi}d\phi \left(r^2\psi^4\sqrt{1+h_\theta^2/r^2}\right)_{r=h(\theta)} \approx 2\pi \int_0^\pi \left(r^2\psi^4\right)_{r=h(\theta)} \sin\theta\,d\theta\,.
\end{equation}
Thus we obtain the area and the horizon mass
\begin{equation}
  \label{eq:70}
  A = 16\pi m_1^2\left( 1+\frac{m_2}{2d}\right)^2 + \mathcal{O}(1/d^4) \Rightarrow M_1 = m_1\left(1+\frac{m_2}{2d}\right) + \mathcal{O}(1/d^4)\,.
\end{equation}
The corresponding expression for $M_2$ is obtained by interchanging
$m_1$ and $m_2$.

The intrinsic metric on the marginal surface is of the form given in
Eq.~(\ref{eq:12}) with $f$ and the coordinate $\zeta$ given as in
Eq.~(\ref{eq:68}).  A straightforward calculation then leads to 
\begin{equation}
  \label{eq:72}
  \R = -\frac{1}{R^2}\frac{d^2f}{d\zeta^2} = \frac{1}{2M_1^2} + \frac{M_2}{4d^3}P_2(\zeta) + \mathcal{O}(1/d^4)\,. 
\end{equation}
We have expressed the result in terms of the physical horizon masses
$M_1$ and $M_2$ rather than the bare parameters $m_1$ and $m_2$. 

It is then interesting to compare this with the work of Damour \&
Lecian \cite{Damour:2009va}.  This result is to be compared with
Eq.~(32) of \cite{Damour:2009va} which can be written in our notation
as
\begin{equation}
  \label{eq:61}
  \R = \frac{1}{2M_1^2}+ \frac{4M_2}{d^3}P_2(\zeta) +\ldots\,.  
\end{equation}
This disagrees with the corresponding term of Eq.~(\ref{eq:72})
containing $P_2(\zeta)$; the Brill-Lindquist black hole is less
distorted.  As discussed at towards the end of the introduction, a
disagreement is not entirely surprising since the Brill-Lindquist data
can be different from the Weyl ansatz used in \cite{Damour:2009va}.
Recent work by Landry and Poisson \cite{Landry:2014jka} reproduces the
results of Damour and Lecian.  However it uses a different formalism,
and we haven't carried out a detailed comparison just yet.

\subsubsection{Incorporating spin effects}

For the case of a spinning black hole with a binary companion, new
terms appear in the higher orders of $J$ and $1/d$ starting with
$\mathcal{O}(J^2/d^2)$.  We write $h(\theta) = h^{BL}(\theta) +
\widetilde{h}(\theta)$, where $h^{BL}$ denotes the Brill-Lindquist
result of Eq.~(\ref{eq:51}).  Then, it can be shown that
\begin{eqnarray}
  \widetilde{h}(\theta) &=& -\frac{3m_2J^2}{200m_1^2d^2}\left(P_1(\cos\theta)-\frac{(4657-6720\ln[2])}{13}P_3(\cos\theta)\right) \nonumber \\
  && - \frac{m_2J^2}{16m_1^2d^3}\left(\frac{47m_1}{70} - \frac{21m_2}{25}P_1(\cos\theta)+\frac{(32567-47040\ln[2])m_1}{343}P_2(\cos\theta) \right. \nonumber \\
  && \left. + \frac{21(4657-6720\ln[2])m_2}{325}P_3(\cos\theta) - \frac{3(391259-564480\ln[2])m_1}{3430}P_4(\cos\theta)\right) \nonumber \\
  \label{eq:40}
  && + \mathcal{O}(J^4) + \mathcal{O} \left( \frac{1}{d^4} \right)\,.
\end{eqnarray}
Thus, unlike the case of a single spinning Bowen-York black hole
studied earlier, the horizon is no longer reflection symmetric, and
the location is not independent of $J$.

We can finally compute the physical quantities of the horizon. As
before, we start with the area:
\begin{eqnarray}
  \label{eq:39}
  A &=& 16 \pi m _{1}^2 \bigg[ \left( 1 + \frac{11 J^2}{20 \, m _{1}^4} \right) 
  + \frac{m _{2}}{d} \left( 1 - \frac{11 J^2}{20 \, m _{1}^4} \right) + \frac{m _{2}^2}{d^2} 
  \left( \frac{1}{4} + \frac{33 J^2}{80 \, m _{1}^4} \right) \nonumber \\
  && + \frac{m _{2}}{d^3} \frac{J^2 \left( 39 m _{1}^2 - 55 m _{2}^2 \right)}{200 \, m _{1}^4} \bigg] 
  + \mathcal{O}(J^4) + \mathcal{O} \left( \frac{1}{d^4} \right) \, . 
\end{eqnarray}
The angular momentum is just $J$ and we can thus easily compute the
horizon mass: 
\begin{equation}
  \label{eq:60}
  M_1 = m_1 \left[ 1 + \frac{m_2}{2d} + \frac{2}{5}\frac{J^2}{m_1^4} - \frac{3}{5}\frac{m_2J^2}{m_1^4d} + \frac{3}{5}\frac{m_2^2J^2}{m_1^4d^2} - \frac{1}{2}\frac{m_2^3J^2}{m_1^4d^3} + \frac{39}{400}\frac{m_2J^2}{m_1^2d^3}\right]
\end{equation}
It is easy to check that previous results are recovered for either a
single spinning black hole ($d=\infty$), or for a non-spinning black
hole with a binary companion ($J=0$).

Using the solution above for the conformal factor and the horizon
location, as before, we follow the procedure of computing the
preferred coordinate $\zeta$ and the scalar curvature $\R$.  We shall
not show the intermediate results, but rather just move on to the main
quantities of interest, i.e. the multipole moments.

Apart from the mass and angular momentum, the non-vanishing multipole
moments are (as usual, all results ignore terms of
$\mathcal{O}(1/d^4)$ or $\mathcal{O}(J^3)$ or higher):
\begin{eqnarray}
  \M{2} &\approx& \frac{2}{5}\frac{m_1^5m_2}{d^3} - \frac{2}{25} \frac{J^2}{m_1} + \frac{1}{25}\frac{ J^2}{m_1} \frac{m _{2}}{d} 
  - \frac{1}{50}\frac{ J^2}{m_1} \frac{m _{2}^2}{d^2} +\frac{1}{100}\frac{J^2}{m_1}\frac{m_2^3}{d^3}\nonumber \\
  && + \frac{( - 5294 + 7680 \ln[2])}{100}\frac{m_1m_2}{d^3}J^2  \, , \\
  \M{3} &\approx & \frac{72 m _{1} m _{2} J^2}{35 d^2} (111 - 160 \ln[2]) - \frac{72 m _{1} m _{2}^2 J^2}{35 d^3} (111 - 160 \ln[2])\nonumber \, ,
\end{eqnarray}
\begin{eqnarray}
\J{2} &\approx& - \frac{6 m_{1}^2 m_{2} J}{5 d^2} + \frac{3 m_{1}^2 m_{2}^2 J}{5 d^3}  \\
\J{3} &\approx& - \frac{6 m_{1}^4 m_{2} J}{7 d^3}  \, .
\end{eqnarray}
One might suspect that these results would simplify by using the
physical horizon masses instead of the bare masses $m_1$, $m_2$.  This
is indeed the case.  For $\M{2}$ we get
\begin{equation}
  \label{eq:73}
  \M{2} \approx -\frac{2}{25}M_1^3\chi^2 + \frac{2}{5}\frac{M_1^5M_2}{d^3} + k \frac{M_1^5M_2}{d^3}\chi^2\,.
\end{equation}
Here, as defined earlier, $\chi := J/M_1^2$ and 
\begin{equation}
  \label{eq:74}
  k = \frac{- 5294 + 7680 \ln[2]}{100} - \frac{2}{25} \approx 0.2137 \,.  
\end{equation}
The first two terms of $\M{2}$ have been calculated earlier and the
first effect of spin appears through the coefficient $k$.  Viewing
this as a perturbation of $\M{2}$:
\begin{equation}
  \label{eq:75}
  \frac{\delta\M{2}}{M_1^3} = \frac{2}{5}\frac{M_1^2M_2}{d^3} + k\frac{M_1^2M_2}{d^3}\chi^2 + \ldots\,. 
\end{equation}
Thus, we conclude that the tidal coefficients (defined in
Eq.~(\ref{eq:25})) $\alpha^{(2)}_{210} = 2/5$ and $\alpha^{(2)}_{212}
= k$ characterize the perturbations of the mass quadrupole moment.

Consider now the third mass moment $\M{3}$.  This is rewritten as: 
\begin{equation}
  \label{eq:76}
  \frac{\M{3}}{M_1^4} \approx \frac{72h}{35}\frac{M_1M_2}{d^2}\chi^2 - \frac{108h}{35}\frac{M_1M_2^2}{d^3}\chi^2 - \frac{36h}{35}\frac{M_1^2M_2}{d^3}\chi^2 \ldots
\end{equation}
where $h:= 110-160\ln(2) \approx 0.0965$.  This determines the tidal
coefficients $\alpha^{(3)}_{112}$, $\alpha^{(3)}_{122}$ and
$\alpha^{(3)}_{212}$.

Finally, turning to the angular momentum moments:
\begin{equation}
  \label{eq:77}
  \frac{\delta \J{2}}{M_1^3} \approx -\frac{6}{5}\frac{M_1M_2}{d^2}\chi + \frac{9}{5}\frac{M_1M_2^2}{d^3}\chi + \frac{3}{5}\frac{M_1^2M_2}{d^3}\chi
\end{equation}
and
\begin{equation}
  \label{eq:78}
  \frac{\delta \J{3}}{M_1^4} \approx - \frac{6 M_{1}^2 M_{2} \chi}{7 d^3}  \, .
\end{equation}
This determines the tidal coefficients
\begin{equation}
  \beta_{111}^{(2)} = - \frac{6}{5} \, , \qquad \beta_{121}^{(2)} = \frac{9}{5} \, ,\qquad \beta_{211}^{(2)} = \frac{3}{5}\, , \qquad  \beta_{211}^{(3)} = - \frac{6}{7} \, .
\end{equation}
These coefficients describe the distortion of the spinning black hole
horizon to linear order in the perturbation and up to order $J^2$.

\section{Conclusions}
\label{sec:conclusions}

In this paper we have computed the tidal deformations of the horizon
of a spinning black hole in a binary system using the Bowen-York
initial data.  We have defined a set of dimensionless numbers, the
tidal coefficients, which characterize the deformations.  We have seen
that the effect of the tidal deformations appears already at
$\mathcal{O}(1/d)$ for a spinning black hole.  A number of immediate
generalizations are possible even just for the Bowen-York data for
small angular momentum as considered here. The first is the deviation
from axisymmetry. This includes the case when the spin and the
separation vector are not parallel, including linear momenta for both
black holes, and finally, including spin in the second black hole.
Since we have a flat background metric available to us, all of these
cases can be dealt with.  This will enable us to, for example,
determine circular orbits, find the minimum energy circular orbit, and
compare these results with expectations from post-Newtonian theory.

It would be interesting to compute the tidal coefficients during the
course of a binary black hole numerical simulation.  Assuming that the
horizon geometry can be tracked with sufficient accuracy, this would
give us a more accurate value of the tidal coefficients for the
physical situation that we are interested in, namely two Kerr black
holes orbiting each other and each being distorted by the
gravitational field of the other.  

Most importantly, an important missing piece in the literature on
isolated horizons is the relation between the horizon multipole
moments and the usual field moments at infinity.  There should
similarly be a relation between the tidal coefficients at the horizon
and at infinity.  From the viewpoint of isolated horizons, this is
expected because the horizon geometry (plus the transverse radiation
$\Psi_4$) determines the spacetime in the neighborhood of an isolated
horizon \cite{Ashtekar:2000sz}.  Thus, we can expect that a knowledge
of the tidal coefficients at the horizon should determine the tidal
coefficients for the field moments.  This is well known in Newtonian
theory where the two sets of Love numbers are simply related to each
other.  Landry and Poisson \cite{Landry:2014jka} have determined this
relationship in general relativity for non-spinning neutron stars and
black holes.  It would be interesting to extend these results for
spinning black holes as well.

\section*{Acknowledgments}

We are grateful to Domenico Giulini for valuable comments and discussions.

\appendix

\section{Radial functions for the binary system}
\label{sec:appendix}

We give here the radial functions $u_0\ldots u_4$ of Eq.~(\ref{eq:38})
obtained by solving Eqs.~(\ref{eq:36})-(\ref{eq:36d}):
\begin{eqnarray}
  u_{0} (r) &=& \frac{2 J^2}{5 m _{1}^3 \beta^5} \left( m _{1}^4 + 10 m _{1}^3 r + 40 m _{1}^2 r^2 + 40 m _{1} r^3 + 16 r^4 \right)  \nonumber \\
  && - \frac{1}{d} \frac{m _{2} J^2}{5 m _{1}^3 \beta^6}  \left( 3 m _{1}^5 + 36 m _{1}^4 r + 180 m _{1}^3 r^2 + 480 m _{1}^2 r^3 \right.\nonumber \\
  && \left. + 384 m _{1} r^4 + 128 r^5 \right)\nonumber \\
  && + \frac{1}{d^2} \frac{m _{2}^2 J^2}{5 m _{1}^3 \beta^7} \left( 3 m _{1}^6 + 42 m _{1}^5 r + 252 m _{1}^4 r^2 + 840 m _{1}^3 r^3 \right. \nonumber \\
  && + \left. 1680 m _{1}^2 r^4 + 1120 m _{1} r^5 + 320 r^6 \right)   \nonumber \\
  && - \frac{1}{d^3} \frac{m _{2} J^2}{10 m _{1}^3 \beta^8} \left[ 5 m _{2}^2 \left( m _{1}^7 + 16 m _{1}^6 r + 112 m _{1}^5 r^2 + 448 m _{1}^4 r^3 \right. \right.  \nonumber \\
  && + 1120 m _{1}^3 r^4 + 1792 m _{1}^2 r^5  \left. + 1024 m _{1} r^6 + 256 r^7 \right)  \nonumber \\
  && - m _{1}^2 \beta^2 \left( m _{1}^5 + 12 m _{1}^4 r + 60 m _{1}^3 r^2 + \left. 160 m _{1}^2 r^3 + 240 m _{1} r^4 + 192 r^5 \right)  \right] \, .
\end{eqnarray}
\begin{eqnarray}
  u_{1} (r) &=& \frac{48 m _{2} r^5 J^2}{25 m _{1}^4 \beta^6} \left[ \frac{1}{d^2} \left( 15 m _{1}^2 + 12 m _{1} r + 4 r^2 \right) - \frac{1}{d^3} \frac{4 m _{2} r \left( 21 m _{1}^2 + 14 m _{1} r + 4 r^2 \right)}{\beta} \right] \, .
\end{eqnarray}
\begin{eqnarray}
  u_{2} (r) &=& - \frac{16 J^2 r^2}{5 m _{1} \beta^5} + \frac{8 J^2 m _{2} r^2}{5 m _{1} \beta^6} \frac{1}{d} \left(m _{1}+12r\right)   \nonumber \\
&& - \frac{4 J^2 m _{2}^2 r^2}{5 m _{1} \beta^7} \frac{1}{d^2} \left(m _{1}^2 + 14 m _{1} r + 84 r^2\right)              \nonumber \\
&& + \frac{J^2 m _{2}}{35 m _{1} r^3 \beta^8} \frac{1}{d^3} \bigg( 14 m _{2}^2 r^5 \left( m _{1}^3 + 16 m _{1}^2 r + 112 m _{1} r^2 + 448 r^3 \right) \nonumber \\
&& - 5 m _{1} r \beta^2 \left( 42 m _{1}^6 + 462 m _{1}^5 r + 2072 m _{1}^4 r^2 \right. \nonumber \\
&& + 4788 m _{1}^3 r^3 + 5847 m _{1}^2 r^4 \left. + 3300 m _{1} r^5 + 420 r^6 \right) \nonumber \\
&& - 105 m _{1}^2 \beta^8 \ln \left[ \frac{m _{1}}{\beta} \right] \bigg) \, .
\end{eqnarray}

\begin{eqnarray}
  u_{3} (r) &=& \frac{3 m _{2} J^2}{25 r^4 \beta^6} \frac{1}{d^2} \bigg( 2r \, \big( 15 m _{1}^6 + 165 m _{1}^5 r + 740 m _{1}^4 r^2 + 1710 m _{1}^3 r^3    \nonumber \\
&& + 2088 m _{1}^2 r^4 + 1176 m _{1} r^5 + 140 r^6 \big) \nonumber \\
&& + 15 m _{1} \beta^6 \ln \left[\frac{m _{1}}{\beta} \right] \bigg)    \nonumber \\
&& + \frac{3 m _{2}^2 J^2}{50 r^4 \beta^7} \frac{1}{d^3} \bigg( 2r \, \big( 135 m _{1}^7 + 1755 m _{1}^6 r + 9630 m _{1}^5 r^2 \nonumber \\
&& + 28710 m _{1}^4 r^3 + 49572 m _{1}^3 r^4 + 48168 m _{1}^2 r^5  \nonumber \\
&& + 22408 m _{1} r^6 + 2240 r^7 \big)  \nonumber \\
&& + 135 m _{1} \beta^7 \ln \left[ \frac{m _{1}}{\beta} \right] \bigg) \, .
\end{eqnarray}

\begin{eqnarray}
  u_{4} (r) &=& - \frac{3 m _{2} J^2}{70 r^5 \beta^6} \frac{1}{d^3} \bigg( 2r \, \big( 105 m _{1}^8 + 1155 m _{1}^7 r + 5180 m _{1}^6 r^2   \nonumber \\
&& + 11970 m _ {1}^5 r^3 + 14616 m _{1}^4 r^4 + 8232 m _{1}^3 r^5    \nonumber \\
&& + 960 m _{1}^2 r^6 - 240 m _{1} r^7 + 112 r^8 \big)          \nonumber \\
&& + 105 m _{1}^3 \beta^6 \ln \left[ \frac{m _{1}}{\beta} \right] \bigg) \, .
\end{eqnarray}
We note that $u_1$ does not vanish at spatial infinity. This is
connected to the fact that the approximation of expanding in powers of
$1/d$ is valid only near the first black hole and should not be
expected to be valid away from it.

\section*{References}

\bibliography{ihbib}

\providecommand{\newblock}{}
\begin{thebibliography}{10}
\expandafter\ifx\csname url\endcsname\relax
  \def\url#1{{\tt #1}}\fi
\expandafter\ifx\csname urlprefix\endcsname\relax\def\urlprefix{URL }\fi
\providecommand{\eprint}[2][]{\url{#2}}

\bibitem{Damour:1983}
Damour T 1983 Gravitational radiation and the motion of compact bodies {\em
  Gravitational Radiation\/} ed Deruelle N and Piran T (North-Holland,
  Amsterdam) p~59

\bibitem{Hinderer:2007mb}
Hinderer T 2008 {\em Astrophys.J.\/} {\bf 677} 1216--1220 (\textit{Preprint}
  \eprint{0711.2420})

\bibitem{Flanagan:2007ix}
Flanagan E~E and Hinderer T 2008 {\em Phys.Rev.\/} {\bf D77} 021502
  (\textit{Preprint} \eprint{0709.1915})

\bibitem{Damour:2009vw}
Damour T and Nagar A 2009 {\em Phys.Rev.\/} {\bf D80} 084035 (\textit{Preprint}
  \eprint{0906.0096})

\bibitem{Damour:2012yf}
Damour T, Nagar A and Villain L 2012 {\em Phys.Rev.\/} {\bf D85} 123007
  (\textit{Preprint} \eprint{1203.4352})

\bibitem{Yagi:2013awa}
Yagi K and Yunes N 2013 {\em Phys.Rev.\/} {\bf D88} 023009 (\textit{Preprint}
  \eprint{1303.1528})

\bibitem{Binnington:2009bb}
Binnington T and Poisson E 2009 {\em Phys.Rev.\/} {\bf D80} 084018
  (\textit{Preprint} \eprint{0906.1366})

\bibitem{Damour:2009va}
Damour T and Lecian O~M 2009 {\em Phys.Rev.\/} {\bf D80} 044017
  (\textit{Preprint} \eprint{0906.3003})

\bibitem{Landry:2014jka}
Landry P and Poisson E 2014  (\textit{Preprint} \eprint{1404.6798})

\bibitem{Bowen:1980yu}
Bowen J~M and York Jr J~W 1980 {\em Phys. Rev.\/} {\bf D21} 2047--2056

\bibitem{Garat:2000pn}
Garat A and Price R~H 2000 {\em Phys.Rev.\/} {\bf D61} 124011
  (\textit{Preprint} \eprint{gr-qc/0002013})

\bibitem{Dain:2000hk}
Dain S 2001 {\em Phys.Rev.Lett.\/} {\bf 87} 121102 (\textit{Preprint}
  \eprint{gr-qc/0012023})

\bibitem{Brill:1963yv}
Brill D~R and Lindquist R~W 1963 {\em Phys. Rev.\/} {\bf 131} 471--476

\bibitem{Misner:1960zz}
Misner C~W 1960 {\em Phys. Rev.\/} {\bf 118} 1110--1111

\bibitem{Ashtekar:1998sp}
Ashtekar A, Beetle C and Fairhurst S 1999 {\em Class. Quant. Grav.\/} {\bf 16}
  L1--L7 (\textit{Preprint} \eprint{gr-qc/9812065})

\bibitem{Ashtekar:2000sz}
Ashtekar A {\em et~al.\/} 2000 {\em Phys. Rev. Lett.\/} {\bf 85} 3564--3567
  (\textit{Preprint} \eprint{gr-qc/0006006})

\bibitem{Ashtekar:1999yj}
Ashtekar A, Beetle C and Fairhurst S 2000 {\em Class. Quant. Grav.\/} {\bf 17}
  253--298 (\textit{Preprint} \eprint{gr-qc/9907068})

\bibitem{Ashtekar:2001jb}
Ashtekar A, Beetle C and Lewandowski J 2002 {\em Class. Quant. Grav.\/} {\bf
  19} 1195--1225 (\textit{Preprint} \eprint{gr-qc/0111067})

\bibitem{Ashtekar:2000hw}
Ashtekar A, Fairhurst S and Krishnan B 2000 {\em Phys. Rev.\/} {\bf D62} 104025
  (\textit{Preprint} \eprint{gr-qc/0005083})

\bibitem{Booth:2001gx}
Booth I~S 2001 {\em Class. Quant. Grav.\/} {\bf 18} 4239--4264
  (\textit{Preprint} \eprint{gr-qc/0105009})

\bibitem{Ashtekar:2002ag}
Ashtekar A and Krishnan B 2002 {\em Phys. Rev. Lett.\/} {\bf 89} 261101
  (\textit{Preprint} \eprint{gr-qc/0207080})

\bibitem{Ashtekar:2003hk}
Ashtekar A and Krishnan B 2003 {\em Phys. Rev.\/} {\bf D68} 104030
  (\textit{Preprint} \eprint{gr-qc/0308033})

\bibitem{Hayward:1993wb}
Hayward S 1994 {\em Phys.Rev.\/} {\bf D49} 6467--6474

\bibitem{Hayward:1994yy}
Hayward S~A 1994 {\em Class. Quant. Grav.\/} {\bf 11} 3025--3036
  (\textit{Preprint} \eprint{gr-qc/9406033})

\bibitem{Hayward:2006ss}
Hayward S~A 2006 {\em Phys. Rev.\/} {\bf D74} 104013 (\textit{Preprint}
  \eprint{gr-qc/0609008})

\bibitem{Hayward:2008ti}
Hayward S~A 2008 {Dynamics of black holes} (\textit{Preprint}
  \eprint{0810.0923})

\bibitem{Andersson:2007fh}
Andersson L, Mars M and Simon W 2008 {\em Adv.Theor.Math.Phys.\/} {\bf 12}
  853–888 (\textit{Preprint} \eprint{0704.2889})

\bibitem{Andersson:2005gq}
Andersson L, Mars M and Simon W 2005 {\em Phys.Rev.Lett.\/} {\bf 95} 111102
  (\textit{Preprint} \eprint{gr-qc/0506013})

\bibitem{Dreyer:2002mx}
Dreyer O, Krishnan B, Shoemaker D and Schnetter E 2003 {\em Phys. Rev.\/} {\bf
  D67} 024018 (\textit{Preprint} \eprint{gr-qc/0206008})

\bibitem{Beetle:2008yt}
Beetle C 2008 {Approximate Killing Fields as an Eigenvalue Problem}
  (\textit{Preprint} \eprint{0808.1745})

\bibitem{Beetle:2013zga}
Beetle C and Wilder S 2014 {\em Class.Quant.Grav.\/} {\bf 31} 075009
  (\textit{Preprint} \eprint{1401.0074})

\bibitem{Cook:2007wr}
Cook G~B and Whiting B~F 2007 {\em Phys.Rev.\/} {\bf D76} 041501
  (\textit{Preprint} \eprint{0706.0199})

\bibitem{Lovelace:2008tw}
Lovelace G, Owen R, Pfeiffer H~P and Chu T 2008 {\em Phys.Rev.\/} {\bf D78}
  084017 (\textit{Preprint} \eprint{0805.4192})

\bibitem{Ashtekar:2004gp}
Ashtekar A, Engle J, Pawlowski T and Van Den~Broeck C 2004 {\em Class. Quant.
  Grav.\/} {\bf 21} 2549--2570 (\textit{Preprint} \eprint{gr-qc/0401114})

\bibitem{Ashtekar:2013qta}
Ashtekar A, Campiglia M and Shah S 2013 {\em Phys.Rev.\/} {\bf D88} 064045
  (\textit{Preprint} \eprint{1306.5697})

\bibitem{Schnetter:2006yt}
Schnetter E, Krishnan B and Beyer F 2006 {\em Phys. Rev.\/} {\bf D74} 024028
  (\textit{Preprint} \eprint{gr-qc/0604015})

\bibitem{Chandrasekhar:1985kt}
Chandrasekhar S 1985 {\em {The mathematical theory of black holes}\/} (Oxford
  Classic Texts in the Physical Sciences)

\bibitem{Jaramillo:2011pg}
Jaramillo J~L, Reiris M and Dain S 2011 {\em Phys.Rev.\/} {\bf D84} 121503
  (\textit{Preprint} \eprint{1106.3743})

\bibitem{Misner:1957mt}
Misner C~W and Wheeler J~A 1957 {\em Annals Phys.\/} {\bf 2} 525--603

\bibitem{Gibbons:1972ym}
Gibbons G 1972 {\em Commun.Math.Phys.\/} {\bf 27} 87--102

\bibitem{Brandt:1997tf}
Brandt S and Bruegmann B 1997 {\em Phys.Rev.Lett.\/} {\bf 78} 3606--3609
  (\textit{Preprint} \eprint{gr-qc/9703066})

\bibitem{Cook:1989fb}
Cook G~B and York James~W J 1990 {\em Phys.Rev.\/} {\bf D41} 1077

\bibitem{Ashtekar:2004cn}
Ashtekar A and Krishnan B 2004 {\em Living Rev. Rel.\/} {\bf 7} 10
  (\textit{Preprint} \eprint{gr-qc/0407042})

\bibitem{Dennison:2006nq}
Dennison K~A, Baumgarte T~W and Pfeiffer H~P 2006 {\em Phys.Rev.\/} {\bf D74}
  064016 (\textit{Preprint} \eprint{gr-qc/0606037})

\bibitem{BK-thesis}
Krishnan B 2002 {\em Isolated horizons in numerical relativity\/} Ph.D. thesis
  Pennsylvania State University
  \urlprefix\url{https://etda.libraries.psu.edu/paper/5969/}

\end{thebibliography}
\bibliographystyle{iopart-num}

\end{document}